\documentclass[submission, Proceedings]{SciPost}

\hypersetup{
    colorlinks,
    linkcolor={red!50!black},
    citecolor={blue!50!black},
    urlcolor={blue!80!black}
}

\usepackage[bitstream-charter]{mathdesign}
\usepackage{color}
\newcommand{\tH}{t_{\rm h}}

\DeclareSymbolFont{usualmathcal}{OMS}{cmsy}{m}{n}
\DeclareSymbolFontAlphabet{\mathcal}{usualmathcal}

\begin{document}

\begin{center}{\Large \textbf{
Photoinduced pairing in Mott insulators
}}\end{center}

\begin{center}
Satoshi Ejima\textsuperscript{1,2$\star$}
and 
Holger Fehske\textsuperscript{1,3}
\end{center}

\begin{center}
{\bf 1} Institut f\"ur Physik, Universit\"at Greifswald, 17489 Greifswald, Germany
\\
{\bf 2} Institut f\"ur Softwaretechnologie, Abteilung High-Performance Computing, 
        Deutsches Zentrum f\"ur Luft- und Raumfahrt (DLR), 22529 Hamburg, Germany
\\
{\bf 3} Erlangen National High Performance Computing Center, Friedrich-Alexander-Universit\"at
  Erlangen-N{\"u}rnberg, 91058 Erlangen, Germany
\\
* satoshi.ejima@dlr.de
\end{center}

\begin{center}
\today
\end{center}

\definecolor{palegray}{gray}{0.95}
\begin{center}
\colorbox{palegray}{
  \begin{tabular}{rr}
  \begin{minipage}{0.1\textwidth}
    \includegraphics[width=30mm]{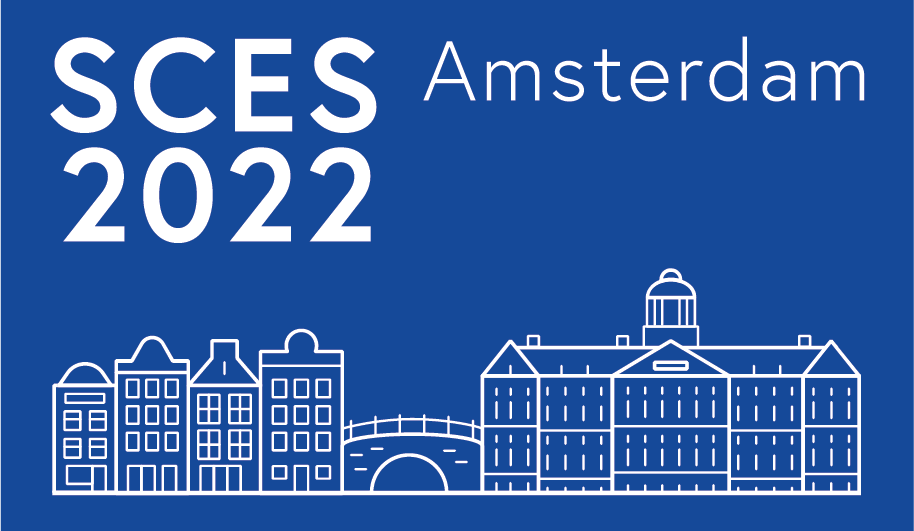}
  \end{minipage}
  &
  \begin{minipage}{0.85\textwidth}
    \begin{center}
    {\it International Conference on Strongly Correlated Electron Systems\\ (SCES 2022)}\\
    {\it Amsterdam, 24-29 July 2022} \\
    \doi{10.21468/SciPostPhysProc.?}\\
    \end{center}
  \end{minipage}
\end{tabular}
}
\end{center}

\section*{Abstract}
{\bf
Utilizing time-evolution techniques in (infinite) matrix-product-state representation, we study the non-equilibrium dynamics of driven Mott insulators and demonstrate photoinduced $\eta$ pairing directly in the thermodynamic limit. Analyzing the time evolution of the corresponding pairing correlations, we determine the optimal laser pump parameters for which long-range $\eta$-pairing becomes dominant after pulse irradiation. The time-dependent photoemission spectra for this optimal pump parameter set show clear signatures of the photoinduced insulator-to-metal phase transition related to the formation of $\eta$ pairs.
}

\vspace{10pt}
\noindent\rule{\textwidth}{1pt}
\tableofcontents\thispagestyle{fancy}
\noindent\rule{\textwidth}{1pt}
\vspace{10pt}

\section{Introduction}
\label{sec:intro}
$\eta$ pairing, proposed first by C. N. Yang in 1989~\cite{Yang89}, gives rise to a pairing-density-wave-like off-diagonal long-range order in the Hubbard model.  While it can be used to construct exact eigenstates of this model, $\eta$ pairing is absent in the Hubbard model's ground state, and therefore has attracted only specific attention, mostly from mathematical point of view. Recently, however, it was pointed out that the $\eta$-pairing state will be enforced by pulse irradiation~\cite{Kaneko19}. The respective enhancement of pairing correlations emerged in time-dependent exact diagonalisations: Calculating all eigenstates as well as pairing correlations for a small cluster and taking the selection rule of $\eta$ pairs into account, Kaneko {\sf et al.} showed that this photoinduced state is related to the $\eta$-pairing state~\cite{Kaneko19}.

Meanwhile, as a result of on-going developments in (time-depenent) density-matrix renormalisation group [(t-)DMRG] technique~\cite{White92,Sch11}, optically driven systems in (quasi-)one-dimen\-sion can be simulated directly in the thermodynamic limit. In doing so, static correlation functions such as $\eta$-pair correlations can be computed by means of the infinite time-evolving block decimation (iTEBD) technique~\cite{iTEBD}, taking advantage of translational invariance in the infinite matrix-product-state (iMPS) representation. Building window sites with so-called infinite boundary conditions (IBC) in the uniform update scheme~\cite{Zauner2015} enables us to simulate non-equilibrium dynamics of excited (quasi-)one-dimensional (1D) systems by a laser electric field~\cite{PhysRevResearch.4.L012012}.

On this basis, in this study, we reexamine the time-evolution of photoinduced $\eta$-pairing, mainly to confirm or put in question previous small cluster results. Thereby we emphasize the importance of using optimal pump pulse parameters. Furthermore, we reconsider the relation between the $\eta$-pairing correlations and the optical spectrum in the small-amplitude regime after pulse irradiation. Finally we prove the photoinduced insulator-to-metal phase transition by simulating time-dependent photoemission spectra of driven Mott insulators.

\section{Model}
\label{sec:model}
Let us consider the 1D half-filled Hubbard model, 
\begin{align}
 \hat{H}=
 -\tH \sum_{j,\sigma}
  \big(
   \hat{c}_{j,\sigma}^\dagger \hat{c}_{j+1,\sigma}^{\phantom{\dagger}}
   +\text{H.c.}
  \big)
  +U\sum_{j}\left(
     \hat{n}_{j,\uparrow}-1/2
    \right)
    \left(
    \hat{n}_{j,\downarrow}-1/2
   \right)\,,
  \label{hubbard}
\end{align}
where  $\tH$ is the nearest-neighbor transfer amplitude and $U$ gives the on-site part of the Coulomb interaction. 
In Eq.~\eqref{hubbard}, $\hat{c}_{j,\sigma}^{\dagger}$ ($\hat{c}_{j,\sigma}^{\phantom{\dagger}}$)
creates (annihilates) a spin-$\sigma$ ($=\uparrow,\downarrow$) electron at Wannier lattice site $j$, 
and  $\hat{n}_{j,\sigma}=\hat{c}_{j,\sigma}^{\dagger} \hat{c}_{j,\sigma}^{\phantom{\dagger}}$. 
In the repulsive case ($U>0$) the model realizes a Mott insulating ground state  with a finite charge gap $\Delta$. 

Exact eigenstates of the Hubbard model can be constructed by means of the operators
$\hat{\eta}^+ = \sum_j(-1)^j \hat{\Delta}_j^\dagger$, $\hat{\eta}^- = (\hat{\eta}^+)^\dagger$, and
$\hat{\eta}^z = \frac{1}{2}\sum_{j}  (\hat{n}_{j,\uparrow}+\hat{n}_{j,\downarrow}-1)$, where $\hat{\Delta}_j^\dagger=\hat{c}_{j,\downarrow}^\dagger \hat{c}_{j,\uparrow}^\dagger$ denotes the singlet pair-creation operator~\cite{Yang89}.
These so-called $\eta$ operators fulfill SU(2) commutation relations  $[\hat{\eta}^+,\hat{\eta}^-]=2\hat{\eta}^z$ and 
$[\hat{\eta}^z,\hat{\eta}^\pm]=\pm\hat{\eta}^\pm$.  Apparently, the Hubbard Hamiltonian~\eqref{hubbard} commutes with 
$\hat{\eta}^2=\tfrac{1}{2}(\hat{\eta}^+\hat{\eta}^-+\hat{\eta}^-\hat{\eta}^+) +  (\hat{\eta}^z)^2$, i.e., $\langle \eta^2 \rangle$ is a conserved quantity. Long-ranged pairing correlations $\langle\hat{\eta}^+_j\hat{\eta}^-_{\ell}\rangle$ develop when the expectation value $\langle \hat{\eta}^2 \rangle$ becomes finite,  but such $\eta$-pairing states cannot be the ground state of the Hubbard model~\cite{Yang89}. Pulse irradiation can establish $\eta$-paired states in Mott insulators however~\cite{Kaneko19}.

To address this issue, we apply a pump pulse with amplitude $A_0$, frequency $\omega_{\rm p}$ and
width  $\sigma_{\rm p}$, centered at time $t_0(>0$):
\begin{equation}
 A(t)=A_0 e^{-(t-t_0)^2/(2\sigma_{\rm{p}}^2)}\cos\left[\omega_{\rm{p}}(t-t_0)\right] \, .
\end{equation}
The external time-dependent electric  field $A(t)$ changes the hopping amplitude by a Peierls phase~\cite{Peierls1933}: 
$\tH \hat{c}_{j,\sigma}^\dagger \hat{c}_{j+1,\sigma}^{\phantom{\dagger}}\to \tH e^{\mathrm{i}A(t)} \hat{c}_{j,\sigma}^\dagger \hat{c}_{j+1,\sigma}^{\phantom{\dagger}}$, i.e., $\hat{H}\to\hat{H}(t)$. As a result, the system being initially in the ground state, is driven out of equilibrium, $|\psi(0)\rangle\to |\psi(t)\rangle$.

\section{Pairing correlations}
\begin{figure}[tb]
 \centering
 \includegraphics[width=0.7\textwidth]{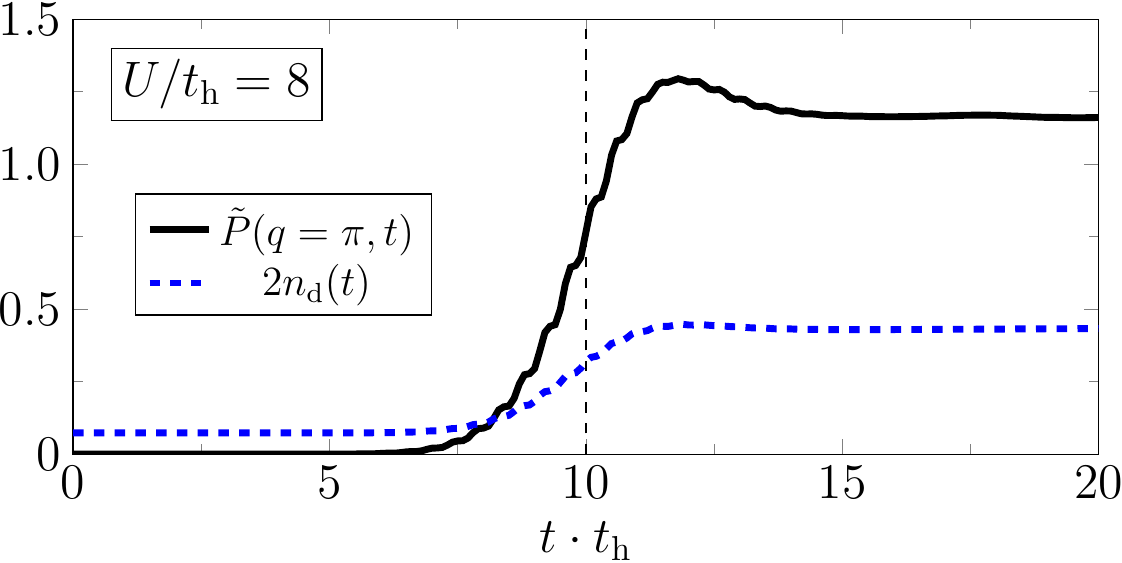}
 \caption{Typical time-evolution process of $\tilde{P}(q=\pi,t)$ and $2n_{\rm d}(t)$ 
 for the photoinduced $\eta$-pairing states in the strong-coupling regime of the driven Hubbard model with $U/\tH=8$  and pump parameters $A_0$=0.4, $\omega_{\rm p}/\tH=7.0$, $\sigma_{\rm p}=2\tH^{-1}$ and $t_0=10\tH^{-1}$.
 The iTEBD data are obtained for bond dimension $\chi=1200$, ensuring a truncation error smaller than $10^{-5}$. For the iTEBD calculations, we employ a second-order Suzuki-Trotter decomposition with time step $0.1\tH^{-1}$ ($0.01\tH^{-1}$).}
 \label{Pq0t}
\end{figure}

The $\eta$-pairing state can be detected evaluating the time evolution of the pair-correlation function
\begin{align}
P(r,t)=\frac{1}{L}\sum_j \langle \psi(t)|\hat{\Delta}^\dagger_{j+r}\hat{\Delta}_j^{\phantom{\dagger}}+{\rm H.c.}) |\psi(t)\rangle\,
\end{align}
and its Fourier transform $\tilde{P}(q,t)=\sum_r e^{\mathrm{i}q r} P(r,t)$. As found in Refs.~\cite{Kaneko19,SCES19} for small clusters, $\tilde{P}(\pi,t)$ is enhanced after pulse irradiation, indicating  the formation of an $\eta$-pairing state. By means of iTEBD this is confirmed directly in the thermodynamic limit which is demonstrated in Fig.~\ref{Pq0t} for a pump with $A_0=0.4$, $\omega_{\rm p}/\tH=7.0$ and $\sigma_{\rm p}=2\tH^{-1}$ centered at $t_0=10\tH^{-1}$.
$\tilde{P}(\pi,t)$ shows a clear response to pulse irradiation and is strengthened as the system progresses in time until saturation is reached. Obviously, the  nonlocal contributions have a stronger impact on $\tilde{P}(\pi,t)$ than 
the double occupancy $n_{\rm d}(t)=(1/L)\sum_j\langle\psi(t)|\hat{n}_{j,\uparrow}\hat{n}_{j,\downarrow}|\psi(t)\rangle$ 
[note that $P(r=0,t)=2n_{\rm d}(t)$, where $n_{\rm d}(0)>0$ for the finite $U$ values considered].

The enhancement process of $\eta$-pairing can be described as follows~\cite{Kaneko19}:
The initial state before pulse irradiation is the ground state of the Hubbard chain with  
$|\eta=0, \eta^z=0\rangle$, which is consistent with the numerical finding: $\tilde{P}(0,t=0)\simeq 0$ (see  Fig.~\ref{Pq0t}).
Turning on the pump pulse, the Hamiltonian does not commute with the $\eta$-operators anymore,
\begin{equation}
 [\hat{H}(t),\eta^+]=[\hat{H},\eta^+] \cos[A(t)]+\sum_k F(k,t) \hat{c}_{\pi-k,\downarrow}^\dagger \hat{c}_{k,\uparrow}^\dagger\,,
\end{equation}
where $F(k,t)=4\tH\sin[A(t)]\sin k$. This alters the initial state to a state with a finite expectation value 
$\langle \hat{\eta}^2\rangle$. Even though the commutation relation is recovered for $t\gg t_0$, i.e., 
$[\hat{H}(t),\hat{\eta}^+]\to[\hat{H},\hat{\eta}^+]$ [since $A(t)\to 0$], $|\psi(t)\rangle$ now includes components of $|\eta>0, \eta^z=0\rangle$ leading to the enhancement of $\tilde{P}(\pi,t)$, see Fig.~\ref{Pq0t} for $t>t_0$.

\begin{figure}[t]
 \centering
 \includegraphics[width=0.95\textwidth]{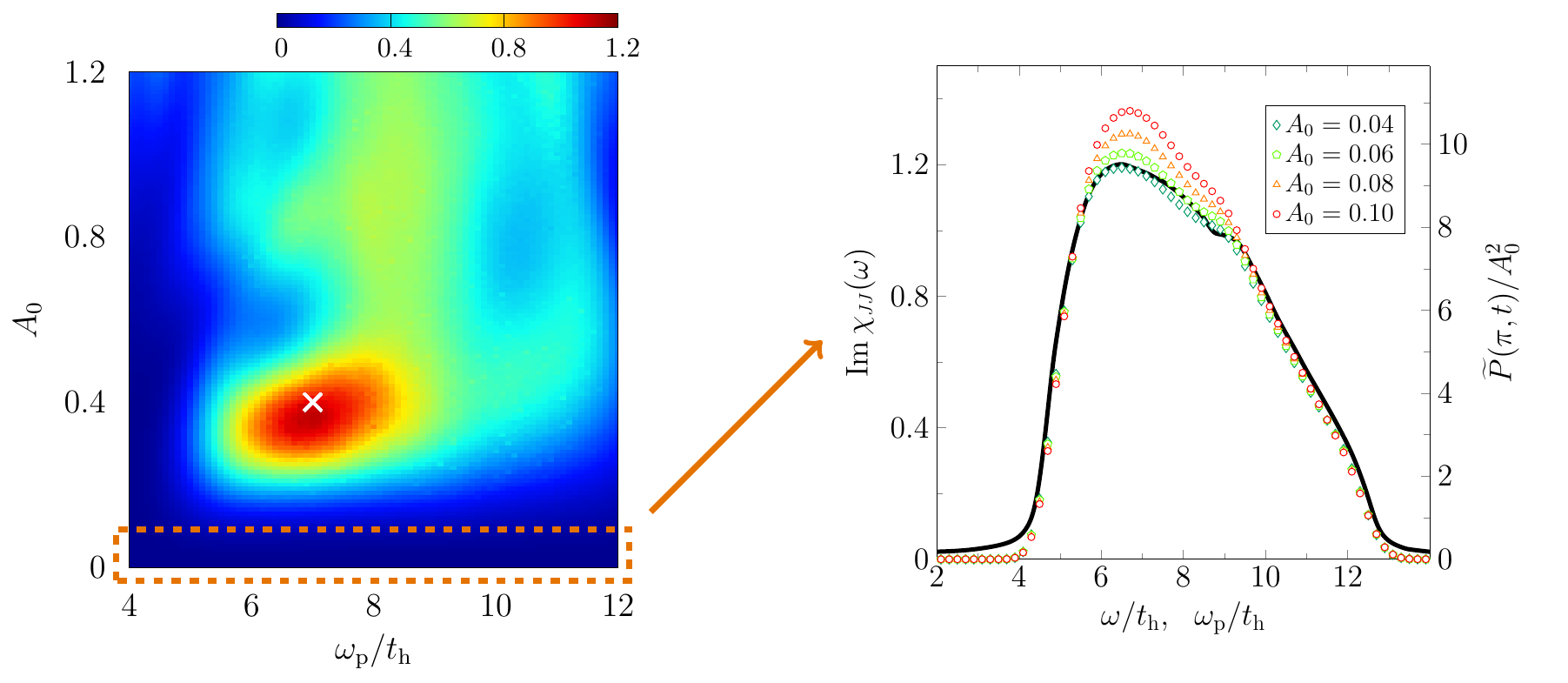}
 \caption{
 (a): Contour plots of $\tilde{P}(q=\pi, t)$ in the $\omega_{\rm p}$-$A_0$ plane at $t=15\tH^{-1}$. Again $U/\tH=8$, and the pump is parametrized by $\sigma_{\rm p}=2\tH^{-1}$ at $t_0=10\tH^{-1}$.
 (b): $\tilde{P}(\pi,t)$ at $t=15\tH^{-1}$ in the small-$A_0$ area enclosed by the dashed square in panel (a). Dividing by $A_0^2$, data can be rescaled to ${\rm Im} \chi(\omega)$ (black line), where ${\rm Im} \chi(\omega)$ is the imaginary part of the optical spectrum $\chi_{JJ}(\omega)$. 
 }
 \label{contour}
\end{figure}
Note that the enhancement of $\eta$ pairing after pulse irradiation depends, however, strongly on the pump pulse parameters. The optimal parameter set for inducing $\eta$-pairing states can be determined examining the $A_0$ and $\omega_{\rm p}$ dependences of $\tilde{P}(\pi,t)$ by iTEBD. Figure~\ref{contour}(a) shows the contour plot of $\tilde{P}(\pi,t)$ after pulse irradiation ($t=15\tH^{-1}$). We find a single maximum around $A_0\approx0.4$ and $\omega_{\rm  p}/\tH\approx 7.0$ (marked by the ``$\times$" symbol), instead of the stripe structure observed in the finite-system ($L=14$) exact diagonalisation (ED) simulations~\cite{Kaneko19}.

Another notable results of previous ED calculations~\cite{Kaneko19} was that the peak structure of $\tilde{P}(\pi,t)$ as a function of $\omega_{\rm p}$ for small $A_0$ is essentially the same as those of the ground-state optical spectrum,
\begin{equation}
 \chi_{JJ}(\omega>0)=-\frac{1}{L}\langle \psi_0| \hat{J}\frac{1}{E_0-\hat{H}+\hbar\omega+\mathrm{i}\eta_{\rm L}}\hat{J}|\psi_0\rangle\,,
 \label{chiii}
\end{equation}
where $|\psi_0\rangle$ is the ground state having energy $E_0$ and  Lorentzian width $\eta_{\rm L}$.  
In \eqref{chiii}, the Hubbard-model charge-current operator is  $\hat{J}=\mathrm{i}\tH\sum_{j,\sigma}(\hat{c}_{j,\sigma}^\dagger\hat{c}_{j+1,\sigma}^{\phantom{\dagger}}-\hat{c}_{j+1,\sigma}^\dagger\hat{c}_{j,\sigma}^{\phantom{\dagger}})$.

Figure~\ref{contour}(b) compares the iTEBD data, obtained for $\tilde{P}(\pi,t)$ at various small $A_0$ and $t=15\tH^{-1}$,  with the t-DMRG results for $\chi_{JJ}(\omega)$ (using $\eta_{\rm L}/\tH=0.2$), in dependence on $\omega_{\rm p}$ respectively $\omega$. Most notably, $\tilde{P}(\pi,t)$ divided by $A_0^2$ scales to the imaginary part of the optical spectrum ${\rm Im}\chi(\omega)$. This can be understood as follows: The hopping term including the Peierls phase can be divided into kinetic and current operators as 
\begin{equation}
 -\tH \sum_{j,\sigma}
  \big(
   e^{\mathrm{i}A(t)}\hat{c}_{j,\sigma}^\dagger \hat{c}_{j+1,\sigma}^{\phantom{\dagger}}
   +\text{H.c.}
  \big)
  =\hat{K}\cos[A(t)]+\hat{J}\sin[A(t)]\,,
 \label{KJ}
\end{equation}
where $\hat{K}=-\tH\sum_{j,\sigma}(\hat{c}_{j,\sigma}^\dagger \hat{c}_{j+1,\sigma}^{\phantom{\dagger}}+\text{H.c.})$.
For small $A_0$ and large $t$, the second term in Eq.~\eqref{KJ} can be approximated by $\hat{J}A_0$, yielding a significant contribution of $A_0^2$ to the pair correlations. 
Needless to say that the finite-size effects are eliminated by simulating the pair correlations directly in thermodynamic limit by iTEBD, leading to the single-peak structure in Fig.~\ref{contour}(b), in strong contrast to the multiple-peak structure observed in the ED calculations~\cite{Kaneko19}.

\section{Non-equilibrium dynamics}
\label{seq:noneq-dyn} 
\begin{figure}[tb]
 \centering
 \includegraphics[width=0.95\textwidth]{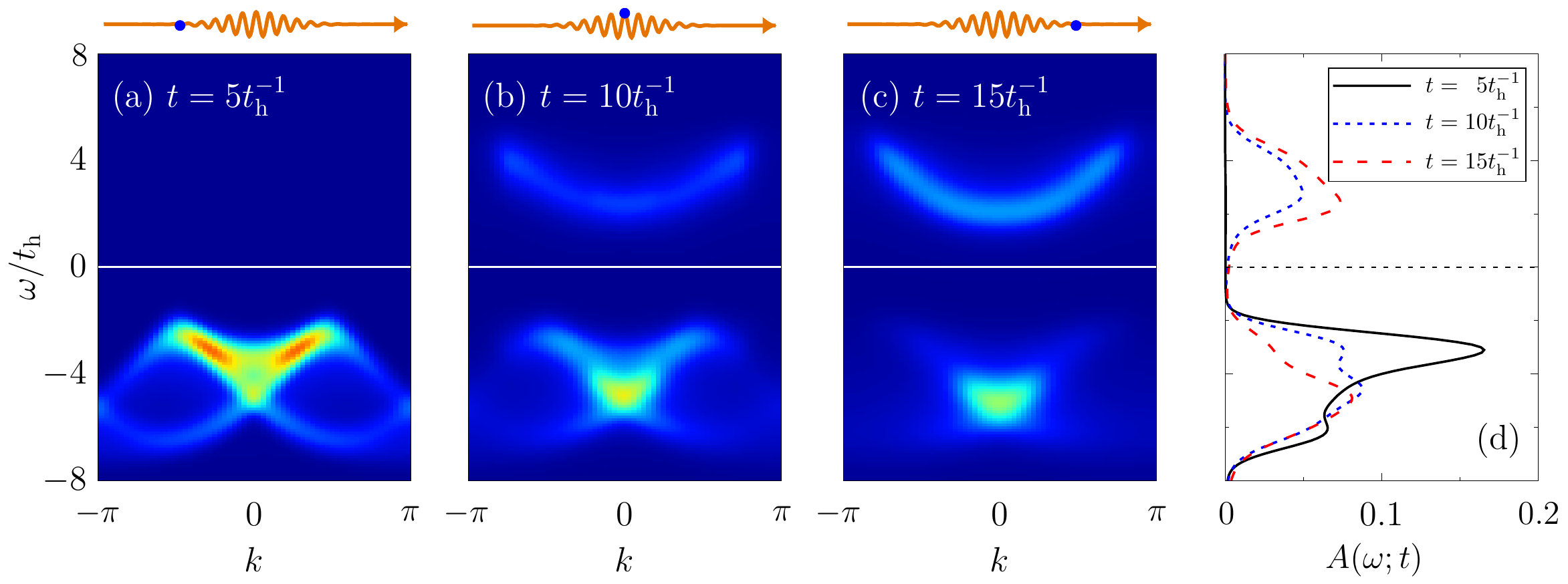}
 \caption{
 Snapshots of the photoemission spectra $A(k,\omega;t)$ indicating photoinduced $\eta$-pairing during the pump at times $t=5\tH^{-1}$ (a), $10\tH^{-1}$ (b) and $15\tH^{-1}$ (c). The pump is parametrized by $A_0=0.4$, $\omega_{\rm p}/\tH=7.0$ [see `$\times$'-symbol in Fig.~\ref{contour}(a)], and $\sigma_{\rm p}=2\tH^{-1}$ at $t_0=10\tH^{-1}$. The transient integrated density of states $A(\omega;t)$ obtained from the data of panels (a)-(c) is depicted in panel (d).  All data are obtained by the (i)TEBD technique with IBC for the 1D half-filled Hubbard model with $U/\tH=8$. Note that the time cutoff in the simulation of time-dependent correlation functions is $T=5\tH^{-1}$,  i.e., the integration in Eq.~\ref{noneq-dyn} extends only over the interval $-T\leq\tau_1,\tau_2 \leq T$. As a compromise between time and frequency resolutions we have chosen a probe pulse width $\sigma_{\rm pr}=2\tH^{-1}$.}
 \label{NoneqDyn}
\end{figure}
We now analyze the non-equilibrium photoemission spectra 
$A(k,\omega;t)=\sum_{\sigma=\uparrow,\downarrow}A_\sigma(k,\omega;t)$
for the optimal pump parameter set marked by the ``$\times$"-symbol in Fig.~\ref{contour}(a).
To explore the system dynamics in a non-equilibrium situation, time-dependent spectral functions of the form~\cite{PhysRevLett.102.136401}
\begin{align}
A_\sigma(k,\omega;t) = \sum_{r}e^{-\mathrm{i}kr} \int_{-\infty}^{\infty}\int_{-\infty}^{\infty} d\tau_1 d\tau_2\, f(\tau_1,\tau_2;\omega)
\cdot C_\sigma(r,\tau_1,\tau_2; t) 
\label{noneq-dyn}
\end{align}
are of interest. Here, the non-equilibrium two-point correlator 
\begin{align}
C_\sigma(r,\tau_1,\tau_2; t)=\langle\phi(t)|\hat{c}_{j+r,\sigma}^\dagger(\tau_1;t)\hat{c}_{j,\sigma}(\tau_2;t)|\phi(t)\rangle
\end{align}
is defined relative to $t$,   and 
\begin{align}
f(\tau_1,\tau_2;\omega)=e^{{\mathrm i}\omega(\tau_1-\tau_2)}g(\tau_1)g(\tau_2)\,,\;\;
g(\tau)=\exp[-\tau^2/2\sigma_{\rm pr}^2]/\sqrt{2\pi}\sigma_{\rm pr}
\end{align}
  specify the shape of the probe pulse, e.g., in a time-dependent photoemission spectroscopy experiment. How numerically  simulate two-time-dependent quantities such as $C_\sigma(r,\tau_1,\tau_2; t)$ has been explained in detail in Ref.~\cite{PhysRevResearch.4.L012012} [see paragraphs below Eq.~(1)].

Figure~\ref{NoneqDyn} displays our (i)TEBD results for the 1D half-filled Hubbard model in the strong-coupling regime ($U/\tH=8$). Before pump irradiation the state is a Mott insulator with a noticable single-particle gap, see Fig.~\ref{NoneqDyn}(a) for $t=5\tH^{-1}$. In the midst of the pump ($t=10\tH^{-1}$), an extra dispersion above Fermi energy ($\omega>E_{\rm F}$) appears and persists afterwards [Fig.~\ref{NoneqDyn}(c)]. 

Evaluating the integrated density of states 
\begin{equation}
 A(\omega;t)=\frac{1}{L}\sum_k A(k,\omega;t)\,,
\end{equation}
we see more clearly how the spectral weight is shifted from $\omega<E_{\rm F}$ to $\omega>E_{\rm F}$ due to the photoinduced $\eta$-pairing. Figure~\ref{NoneqDyn}(d) gives $A(\omega;t)$ for the photoinduced $\eta$-pairing state.
Obviously, the spectral weight for $\omega>E_{\rm F}$ increases distinctly over time, indicating a photoinduced phase transition from a Mott insulator to a metallic $\eta$-pairing state. This photoinduced insulator-to-metal transition should be observed in time- and angle-resolved photoemission spectroscopy, when the pure Hubbard model is realized experimentally, e.g., in optical lattices.
We note that the photoinduced phase transition cannot be observed by simulating the time-dependent photoemission spectra with not-optimized pump-pulse parameters, see Ref.~\cite{PhysRevResearch.4.L012012}.

\section{Conclusions}
To summarize, combining tensor-network algorithms with infinite time-evolving block decimation techniques, we revisited the problem of photoinducing $\eta$-pairing states in the one-dimensional Hubbard model at half band filling. This 
allowed us to prove the enhancement of the pairing correlations directly in the thermodynamic limit. We also
determined the optimal pump-pulse parameter set that maximizes the $\eta$-pairing tendency. An $\eta$-pairing related Mott insulator to metal transition could be extracted from the time-dependent photo\-emission spectrum. 

We wish to stress that the numerical approach presented here can be applied to simulate the non-equilibrium dynamics of any (quasi-)one-dimensional translational-invariant system in entire ranges of interacting and driving parameters. For example, the photoinduced metallization of excitonic insulators was demonstrated quite recently in accordance with 
time- and  angle-resolved photoemission spectroscopy experiments on Ta$_2$NiSe$_5$~\cite{PhysRevB.105.245126,Okazaki2018}.

\section*{Acknowledgements}
The iTEBD simulations were performed using the ITensor library~\cite{ITensor}.
\paragraph{Funding information}
S.E. was supported by Deutsche Forschungsgemeinschaft through proj\-ect EJ 7/2-1.

\nolinenumbers

\end{document}